\documentclass[twocolumn, showpacs,preprintnumbers,amsmath,amssymb]{revtex4}

\usepackage[francais]{babel}
\usepackage[latin1]{inputenc}  
\usepackage[T1]{fontenc}       
\usepackage{graphicx,graphics,amssymb}
\usepackage[dvips]{epsfig}
\DeclareGraphicsExtensions{.jpg,.pdf,.mps,.png}

\begin{document}


\title{Some aspects of simulation and realization \\of an optical reference cavity}

\author{Didier Guyomarc'h}
\author{Ga\"etan Hagel}
\author{Cédric Zumsteg}
\author{Martina Knoop}  \email{Martina.Knoop@univ-provence.fr}

\affiliation{Physique des Interactions Ioniques et Mol\'eculaires
(CNRS UMR 6633), Universit\'e de Provence, Centre de Saint
J\'er\^ome, Case C21, 13397 Marseille Cedex 20, France}%

\date{\today}

\begin{abstract}
The interrogation of an ultra-narrow clock transition of a single trapped  ion for optical frequency metrology requires a laser  stabilized to a couple of Hz per second with a linewidth of the same order of magnitude. Today, lasers in the visible have reached the Hz-range in frequency stability, if locked onto a high-finesse, ultra-stable reference cavity. Vertical mounting of the reference cavity can reduce its sensitivity to vibrations as described in \cite{notcutt05}. We have designed a comparable vertical cavity  with an overall length of 150 mm resulting in a Free Spectral Range of 1GHz. Optimisation of the cavity design has been carried out with a Finite-Elements Method, leading to expected relative length variations below 10$^{-14}$ under the influence of gravity acceleration (1 $g$). The  variation of different geometric parameters has been studied. The analysis of the different noise sources shows that, for a regime superior to  a tenth of a hertz, the fast linewidth of the laser will not be limited by the cavity characteristics.

\end{abstract}

\pacs{ 42.60.Da Resonators, cavities, amplifiers, arrays, and rings, 42.62. Laser applications, 07.05.Fb Design of experiments}
\maketitle

\section{Vibration insensitive optical reference cavity}
Frequency metrology in the optical domain is made by locking a spectrally pure laser source to the ultra-narrow transition of an ensemble of atoms or a single trapped ion. The recording time of the error signal which is fed back onto the laser, which serves as local oscillator, depends on the interrogated system, it can be of the order of a second for single ions. For all times shorter than that this laser should be stable to better than one atomic linewidth which is of the order of one Hertz for the various systems.

Extremely well isolated, high-finesse optical cavities are used as frequency references to assure the stability of visible lasers for timescales up to one second, they also allow to reduce the laser's  linewidth to the order of the hertz \cite{young99}. Recently, new geometries and mounting configurations have been proposed for these cavities in order to reduce vibrational effects which tend to deform the cavity and thus can shift its resonance frequencies \cite{notcutt05,nazarova06,webster07,millo09}.

In this work, we consider a vertically mounted cavity. Minimal length variations of the cavity can  be achieved by choosing a tapered form for the cavity spacer. The gain with respect to a horizontally mounted, cylindric cavity can be as large as two orders of magnitude and with less stringent machining conditions  \cite{notcutt05,chen06}. Vibrationally insensitive mounting is made by supporting the cavity near its midplane. As a matter of fact, under upward acceleration the upper part of the cavity will be compressed while the lower part tends to expand. An ideally supported cavity will show a perfect compensation of both effects.
Optimisation of the geometry and appropriate choice of the supporting points can be made by numerical simulation of the cavity using a finite element analysis (FEA).

Here, we present details of the FEA, which has been  carried out in a similar way to the one described in \cite{chen06}. In particular, Chen et al. have also demonstrated that the simulation in the range below some tens of kHz can be made by a static approach, as the wavelength of the excited eigenfrequencies is estimated to be an order of magnitude larger than the dimensions of the cavity. In the present work,  we have investigated the influence of different aspects of the spacer's geometry, as there are the variation of the cone angle, the position and the angle of the evacuation bores,  as well as the influence of chamfers.
Furthermore, we have estimated the overall thermal noise of the designed cavity, and derived its limit frequencies fluctuations at a time scale of one second. Different noise sources, as well as limitations in the stabilization method,  will determine the final linewidth of the laser locked onto this cavity. In the present work, we concentrate on contributions which are due to the cavity's mechanical or thermal properties only. This gives a lower limit for the estimate of the final laser linewidth, with the goal to design a cavity which will be not the limiting factor in the final laser stabilisation stage.

Our application is the interrogation of a single Calcium ion stored in a miniature trap for metrological applications \cite{champenois04}.  Excitation of the ultra-narrow electrical quadrupole transition at 729~nm has been made by a broad-area laser in an extended cavity set-up \cite{houssin03,hagel05}, before using the actual set-up of a lab-built titanium-sapphire laser at 729 nm which will be locked by the Pound-Drever-Hall method \cite{drever83} onto a high-finesse ($\mathcal{F} \simeq 10^5$), fixed-length optical reference cavity. We aim an inherent stability of better than  1 Hz/$\sqrt Hz$ for the clock laser. For practical reasons we have chosen a free spectral range (FSR) of 1~GHz, resulting in the overall length of the cavity of 150~mm. A length fluctuation of 10$^{-9}$, or 0.15~nm on the optical cavity corresponds to a frequency shift of one Hz. To guarantee a high degree of mechanical as well as thermal stability the cavity spacer is  made of ultra-low expansion glass (Corning ULE$^{\circledR}$),  whereas the mirror substrates are from fused silica \cite{numata04}. The optimized design of this reference cavity is the subject of this paper.

\section{Finite element simulation}

\begin{figure}
\includegraphics[width=0.48\textwidth]{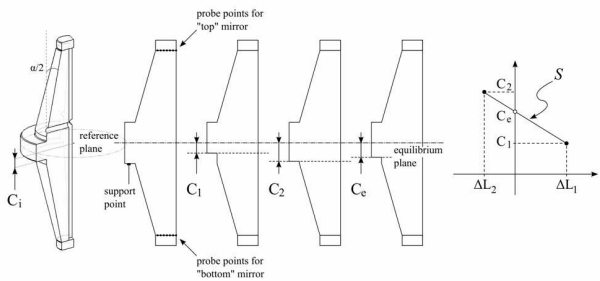}
\caption{Typical geometry of the simulated cavity. Only a third of the spacer is modeled, the simulations take into account the symmetry of the spacer. For detailed explanations see text} \label{fig:cavgeom}\end{figure}

We have chosen the vertically mounted, tapered cavity geometry which  has been shown to be vibrationally insensitive \cite{notcutt05,chen06}. The geometry of the designed cavity is shown in Figure~\ref{fig:cavgeom}, it has been machined from a single piece of Corning ULE$^{\circledR}$.  The mirrors are optically contacted to the spacer, they have a diameter of 25.4 mm and a thickness of 6 mm. The FEA takes into account two plane fused silica mirrors, even though one of them has a radius of curvature of 1~m, the minimal geometrical difference is negligible. The central bore has a diameter of 10~mm, giving ample space for the cavities' TEM00 mode which has a calculated waist diameter on the mirrors below 0.5~mm. Three bore holes in the upper part of the cavity allow evacuation of the inner part for the cavity which is destined to be mounted in a vacuum vessel. The cavity is held vertically by three ceramic posts which support the central collar. The aim is to support the cavity in a plane which corresponds to the vibrational "neutral fiber" of the piece, and to therefore minimize its sensitivity to vertical vibrations.


FEA has been used to optimize the geometry of the ULE spacer as well as to exactly determine the position of the equilibrium support plane.  This latter has been determined in a two-step procedure. In a first phase, simulations of the cavity geometry have been carried out to determine the optimal shape and dimensions, for a fixed  overall length of 150 mm. The cavity has then been machined, and on the finished spacer all dimensions have been measured interferometrically to a hundredth of a millimeter.  In a second step, the cavity has  been updated numerically with the measured values, and the FEA carried out once again to precisely determine the equilibrium plane of support.

All finite element simulations have been carried out using the Cosmos module of the Solidworks software \cite{solidworks08} on a personal computer.  The present analysis fixes the plane of support and calculates the deformation of the cavity under vertical  acceleration of 1 $g$.  This acceleration is assumed uniform and constant in the relevant frequency range up to 10~kHz. Chen et al. \cite{chen06} have shown that for a tapered cavity spacer the influence of horizontal acceleration is at least two orders of magnitude lower than vertical acceleration. A grid of probe points on crucial parts of the cavity (essentially on the mirror surfaces, see figure \ref{fig:cavgeom}) has been defined. Their positional deviation allows to measure the deformations of the cavity spacer in a quantitative way.

To determine the optimal support plane $C_e$ with precision, two points $C_1, C_2$ (see figure \ref{fig:cavgeom}) sufficiently apart are chosen, then the respective length variation of the spacer is calculated. The results are linearly interpolated to determine the equilibrium support plane $C_e$, corresponding to a zero theoretical length variation.  Throughout simulations, the collected fractional length variations  are  inferior to $3\times10^{-12}$ (down to values of $5\times10^{-13}$). In terms of vertical acceleration sensitivity this corresponds to a domain below 1.2 kHz/$g$ at 729~nm.

\section{Simulation results}
A first series of simulations has determined an optimized tapered cavity with a length of 150~mm, and an outside cone angle $\alpha/2$ of 14 degree (see Figure \ref{fig:cavgeom}),  we will use this geometry as a starting point in the following.

Our aim is to investigate the dependence of the cavity geometry on different geometrical aspects, in order to choose a cavity design with a reduced sensitivity to machining uncertainties.
For a quantitative study of the sensitivity of the geometry to the variation of a given geometric parameter, a parameter $\mathcal{S}$ has been defined to be  $\mathcal{S} = \left| (\Delta L_2 - \Delta L_1)/(C_{2} - C_{1})\right|$, where small changes in the height of the probe points $C_i$ give rise to length variations of the spacer $\Delta L_i$.  We have chosen to vary  $C_i$ for  $\pm 0.01$ mm and $\pm 0.1$ mm.


In order to optimize computing ressources, the cavity is cut along its symmetry axes, and only one third of it is simulated.   Computer performance and memory limit the minimum mesh size to 1 mm. Convergence of simulations to a hundredth millimeter can be obtained with mesh sizes as large as 3~mm by manually forcing a tightened mesh distribution insisting on critical aspects as for example the support points, and the evacuation bores.

\begin{figure}[h!]
\includegraphics[width=0.48\textwidth]{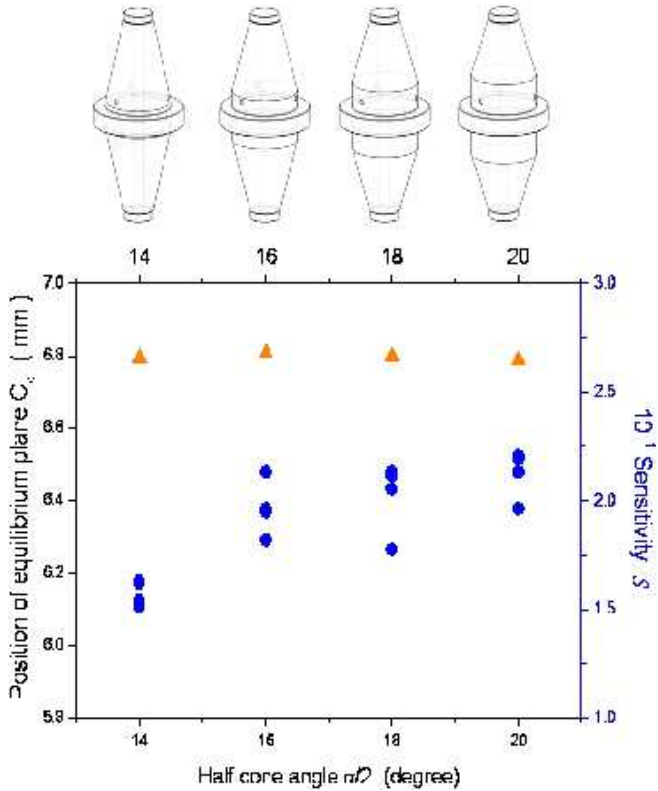}
\caption{(Color online) Variation of the half cone angle of the spacer body. The upper part of the image shows the modification of the spacer's shape with increasing cone angle. The lower part represents the sensitivity $\mathcal{S}$ (bullets, right axis) to the variation of the cone angle and is of the order of 10$^{-11}$, different points represent a different $C_i$ in the simulation. The resulting variation of the height of the support plane (triangles, left axis) is inferior to a tenth of mm.}\label{fig:cone_angle}
\end{figure}

The cone angle $\alpha/2$ is the outside angle of the two halves of the spacer with respect to the central bore. A cylindrical cavity would have an angle of 0$^{\circ}$.
Figure \ref{fig:cone_angle} shows the variation of position of the equilibrium plane as a function of the outside cone angle. Lower cone angles tend to show a smaller sensitivity, a result which is in agreement with the findings of \cite{chen06}, but the overall variation is very small. The differential minimal  displacement of the mirrors can be made to reach values below  10$^{-11}$.

The realisation of the ULE$^{\circledR}$ cavity by a milling machine inevitably creates rough edges on the inflexion points of the complex shape. In glass machining, beveling is often made by hand, making it difficult to estimate the influence of the different chamfers. We have tested the influence of chamfers 0.6$\times $0.6 mm$^2$ on the precision of the proposed simulation. Suppressing chamfers in the simulation of a cavity results in a theoretical displacement of the equilibrium plane of 36 $\mu$m with respect to the cavity where the chamfers are taken into account. This tiny value is of course negligible compared to the machining precision of 0.1 mm.


\begin{figure}[h!]
\includegraphics[width=0.48\textwidth]{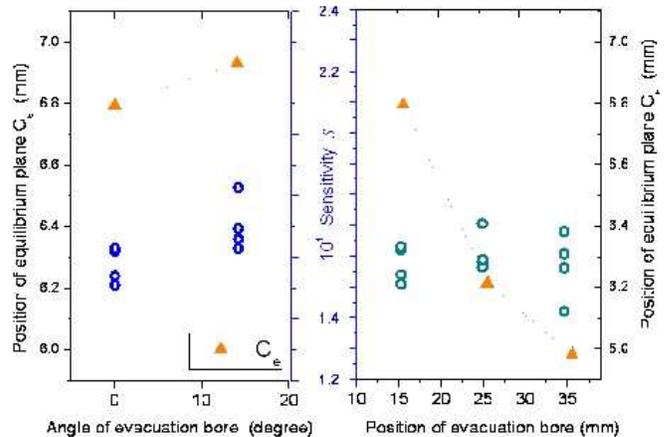}
\caption{(Color online) Variation of the angle (left graph) and position (right graph) of the evacuation bores. Triangles indicate the calculated position of the equilibrium plane on the outside ordinate (the line is a guide to the eye), while the spherical points display the calculated sensitivity for the different $C_i$ on the central ordinate.}\label{fig:evac_bore}
\end{figure}

Figure \ref{fig:evac_bore} shows the effect of position and angle of the three evacuation bores on sensitivity and reference plane. The left part of the figure represents the angle of the holes with respect to the central reference plane, 0 degree corresponds to holes perpendicular to the central bore, an angle of 14 degree corresponds to holes which are drilled in a right angle to the outside of the cone. The observed sensitivity is of the same order of magnitude for both configurations. This is also true for the variation of the height of the evacuation bore (with 0 angle). Nevertheless, it is evident that the absolute value of the equilibrium plane $C_e$ varies for these different geometries.

\vspace{10mm}
In conclusion, among the studied geometric parameters, the variation of the cone angle is the most critical parameter during realization of the cavity spacer. The obtained simulations give very precise results, and metrology of the cavity will result in values precise to a hundredth of a mm. However, the machining process of the glass spacer and mirrors is limited to a precision of a tenth of a mm. Moreover, cone angles or angled bores are often realized with less precision.

In order to minimize errors during the machining process, and to be able to achieve the aimed precisions, an optimized realization protocol would first drill all the bores along axes or perpendicular to axes on a cylindrical spacer. The outside angle of the spacer would only be machined in a consecutive step. Precise metrology of the spacer would then allow to calculate the ideal support plane with precision, before milling it in a last step.

\section{Estimation of the limit linewidth}
Considering the cavity only, without a laser-beam introduced, a major part of the noise is due to inherent mechanical thermal fluctuations. We have estimated the thermal noise limit which affects the ultimate response of the designed cavity. Following the study by Numata et al. \cite{numata04}, the configuration to reach minimal thermal noise has been chosen, which consists of an ULE$^{\circledR}$ spacer with fused silica mirror substrates. The thermal noise limit can then be estimated by the sum of the spectral noise density contributions coming from the spacer body ($G_{sp}$), the mirror substrates ($G_{sub}$), and the mirror coatings ($G_c$)

\begin{widetext}
\begin{align}
G_{total}(f) &= G_{sp}(f) + 2 \left[ G_{sub}(f) + G_c(f) \right]\\
&=-\frac{4k_{B}T}{2 \pi f} \left[  \frac{L}{3\pi R^{2}E_{sp}}\:\Phi_{sp} + \frac{2 (1-\sigma^{2})}{\sqrt{\pi} \omega_{0} E_{sub}}\:\Phi_{sub}
+ (1-2\sigma)(1+\sigma) \frac{4 d}{\pi \omega_{0}^2 E_{c}}\:\Phi_{c}
 \right]
 \label{eq:th_noise}
\end{align}
\end{widetext}

where $T$ is the temperature, chosen to be 300~$K$,  $d$ the diameter of the mirror substrate and $\omega_0$ the waist of the cavity mode. The non-cylindrical geometry of the cavity spacer leads us to choose conservative values for the spacer length, $L$=0.15m, and radius  $R$=0.03m. Young's modulus, $E$, and the Poisson ratio, $\sigma$, describe the elastic and compressional behaviour, they depend on the chosen material just as the mechanical loss factor, $\Phi$, independent of frequency. For ULE these values are $E_{ULE}=6.76 \times 10^{10} Pa$, $\sigma_{ULE}=0.17$ and $\Phi_{ULE}= 1/(6 \times 10^4)$, the values for fused silica are slightly different with $E_{FS} = 7.27 \times 10^{10} Pa$, $\sigma_{FS} = 0.16$ and $\Phi_{FS} = 10^{-6}$ \cite{corning03}. The resulting value for the complete cavity is $\sqrt{G_{total}(f)} = 2.5\times 10^{-17} m/\sqrt{Hz}$, where by far the largest contributions come from the mirror coatings and substrates.

We have checked that all other noise sources in the laser stabilisation are negligible to the presented values. Variations of the refractive index of air due to pressure fluctuations  are largely reduced as the cavity is mounted in a vacuum vessel. At a residual pressure of 10$^{-6}$ mbar, they will contribute to the noise spectral density by less than $10^{-23}/\sqrt{Hz}$ ($10^{-8}Hz/\sqrt{Hz}$ fractional). The shot noise spectral density of a photodiode (required for the electronic lock) is below 8$\times10^{-5} Hz/\sqrt{Hz}$ for a collected input power of $P=1$mW and a typical quantum efficiency of 0.9.  And finally, the intensity fluctuations of a  laser  coupled into a cavity of finesse 10$^5$ will give rise to thermal effects in the mirror coatings as well as a radiation pressure shift. While the first effect is negligible small due to very low absorption values, the second term sums up to a frequency fluctuation of 0.05 Hz for 1~mW of laser power fluctuating less than 1\% \cite{bergquist94}.

Assuming that the contributions of equation \ref{eq:th_noise} are the limiting noise factors for the performances of the reference cavity we can use Halford's theorem \cite{halford71}
\begin{equation}
\int\limits^{+\infty}_{\Delta\nu/2} \frac{G_{total}(f)}{f^2} df=\frac{2}{\pi}
\end{equation}
to estimate the limit laser linewidth $\Delta\nu$ from the frequencies' spectral noise density $G_{total}(f)$.
The above-cited values correspond to a limit linewidth of 120 mHz.

Recent works on highly stabilized visible lasers discuss the lock onto ULE$^{\circledR}$ cavities with the aim to work at the material's thermal expansion's inflexion point in order to minimize thermal drifts \cite{alnis08,dube09,millo09}. Active thermal stabilisation of the overall set-up allows to approach this inflexion point with a typical accuracy of about 10~mK.  The resulting thermal drift is of the order of 2-4 Hz/mK. Three additional stages of passive thermal shielding allow to integrate temperature fluctuations on the cavity to reach an stability of the order of 10~$\mu$K per second. The corresponding linewidth fluctuations are thus much smaller than the above discussed thermal noise features due to mechanical loss.

\section{Conclusion}
In this presentation of the design of a vibrational-insensitive, high-finesse optical reference cavity, we have investigated the sensitivity of the cavities' geometry to some conceptual details. With respect to the design of the cavity,  the outside angle of the spacer (determining the shape) is the most sensitive parameter among those studied here. The presented cavity has a thermal noise limit corresponding to a linewidth of $\Delta\nu = 0.12$ Hz, if the cavities' temperature can be stabilized to within 10~mK of its CTE inflexion point. This critical value is often shifted to very low or even negative temperatures for a cavity with fused silica mirrors, making the temperature stabilisation a more complex operation. The addition of an ULE-ring on the mirror substrates is a means to shift the operational temperature into an easily accessible domain \cite{legero09}.

The presented values have to be compared to the spectral linewidth of the cavities'  Airy peaks  of a few kHz. Advanced frequency control methods have their maximum of gain in the spectral bandwidth of the cavity peak, they allow linewidth reductions of more than three orders of magnitude for the realisation of 1Hz-linewidth lasers \cite{drever83,shevy93,bergquist94,notcutt05,nazarova06}. The presented linewidth estimation is only one order of magnitude below these results, demonstrating that the cavity is not the limiting factor. However, this illustrates the importance of the choice of mirror and spacer materials as well as the careful design of the cavity.


\begin{thebibliography}{19}
\expandafter\ifx\csname natexlab\endcsname\relax\def\natexlab#1{#1}\fi
\expandafter\ifx\csname bibnamefont\endcsname\relax
  \def\bibnamefont#1{#1}\fi
\expandafter\ifx\csname bibfnamefont\endcsname\relax
  \def\bibfnamefont#1{#1}\fi
\expandafter\ifx\csname citenamefont\endcsname\relax
  \def\citenamefont#1{#1}\fi
\expandafter\ifx\csname url\endcsname\relax
  \def\url#1{\texttt{#1}}\fi
\expandafter\ifx\csname urlprefix\endcsname\relax\def\urlprefix{URL }\fi
\providecommand{\bibinfo}[2]{#2}
\providecommand{\eprint}[2][]{\url{#2}}

\bibitem[{\citenamefont{Notcutt et~al.}(2005)\citenamefont{Notcutt, Ma, Ye, and
  Hall}}]{notcutt05}
\bibinfo{author}{\bibfnamefont{M.}~\bibnamefont{Notcutt}},
  \bibinfo{author}{\bibfnamefont{L.}~\bibnamefont{Ma}},
  \bibinfo{author}{\bibfnamefont{J.}~\bibnamefont{Ye}}, \bibnamefont{and}
  \bibinfo{author}{\bibfnamefont{J.}~\bibnamefont{Hall}},
  \bibinfo{journal}{Opt. Lett} \textbf{\bibinfo{volume}{30}},
  \bibinfo{pages}{1815} (\bibinfo{year}{2005}).

\bibitem[{\citenamefont{Young et~al.}(1999)\citenamefont{Young, Cruz, Itano,
  and Bergquist}}]{young99}
\bibinfo{author}{\bibfnamefont{B.~C.} \bibnamefont{Young}},
  \bibinfo{author}{\bibfnamefont{F.~C.} \bibnamefont{Cruz}},
  \bibinfo{author}{\bibfnamefont{W.~M.} \bibnamefont{Itano}}, \bibnamefont{and}
  \bibinfo{author}{\bibfnamefont{J.~C.} \bibnamefont{Bergquist}},
  \bibinfo{journal}{Phys. Rev. Lett.} \textbf{\bibinfo{volume}{83}},
  \bibinfo{pages}{3799} (\bibinfo{year}{1999}).

\bibitem[{\citenamefont{Nazarova et~al.}(2006)\citenamefont{Nazarova, Riehle,
  and Sterr}}]{nazarova06}
\bibinfo{author}{\bibfnamefont{T.}~\bibnamefont{Nazarova}},
  \bibinfo{author}{\bibfnamefont{F.}~\bibnamefont{Riehle}}, \bibnamefont{and}
  \bibinfo{author}{\bibfnamefont{U.}~\bibnamefont{Sterr}},
  \bibinfo{journal}{Appl. Phys. B} \textbf{\bibinfo{volume}{83}},
  \bibinfo{pages}{531} (\bibinfo{year}{2006}).

\bibitem[{\citenamefont{Webster et~al.}(2007)\citenamefont{Webster, Oxborrow,
  and Gill}}]{webster07}
\bibinfo{author}{\bibfnamefont{S.}~\bibnamefont{Webster}},
  \bibinfo{author}{\bibfnamefont{M.}~\bibnamefont{Oxborrow}}, \bibnamefont{and}
  \bibinfo{author}{\bibfnamefont{P.}~\bibnamefont{Gill}},
  \bibinfo{journal}{Phys. Rev. A} \textbf{\bibinfo{volume}{75}},
  \bibinfo{pages}{011801(R)} (\bibinfo{year}{2007}).

\bibitem[{\citenamefont{Millo et~al.}(2009)\citenamefont{Millo, {a}es,
  Mandache, Coq, English, Westergaard, Lodewyck, Bize, Lemonde, and
  Santarelli}}]{millo09}
\bibinfo{author}{\bibfnamefont{J.}~\bibnamefont{Millo}},
  \bibinfo{author}{\bibfnamefont{D.~V.~M.} \bibnamefont{{a}es}},
  \bibinfo{author}{\bibfnamefont{C.}~\bibnamefont{Mandache}},
  \bibinfo{author}{\bibfnamefont{Y.~L.} \bibnamefont{Coq}},
  \bibinfo{author}{\bibfnamefont{E.~M.~L.} \bibnamefont{English}},
  \bibinfo{author}{\bibfnamefont{P.~G.} \bibnamefont{Westergaard}},
  \bibinfo{author}{\bibfnamefont{J.}~\bibnamefont{Lodewyck}},
  \bibinfo{author}{\bibfnamefont{S.}~\bibnamefont{Bize}},
  \bibinfo{author}{\bibfnamefont{P.}~\bibnamefont{Lemonde}}, \bibnamefont{and}
  \bibinfo{author}{\bibfnamefont{G.}~\bibnamefont{Santarelli}},
  \bibinfo{journal}{Physical Review A (Atomic, Molecular, and Optical Physics)}
  \textbf{\bibinfo{volume}{79}}, \bibinfo{eid}{053829}
  (pages~\bibinfo{numpages}{7}) (\bibinfo{year}{2009}),
  \urlprefix\url{http://link.aps.org/abstract/PRA/v79/e053829}.

\bibitem[{\citenamefont{Chen et~al.}(2006)\citenamefont{Chen, Hall, Ye, Yang,
  Zang, and Li}}]{chen06}
\bibinfo{author}{\bibfnamefont{L.}~\bibnamefont{Chen}},
  \bibinfo{author}{\bibfnamefont{J.}~\bibnamefont{Hall}},
  \bibinfo{author}{\bibfnamefont{J.}~\bibnamefont{Ye}},
  \bibinfo{author}{\bibfnamefont{T.}~\bibnamefont{Yang}},
  \bibinfo{author}{\bibfnamefont{E.}~\bibnamefont{Zang}}, \bibnamefont{and}
  \bibinfo{author}{\bibfnamefont{T.}~\bibnamefont{Li}}, \bibinfo{journal}{Phys.
  Rev. A} \textbf{\bibinfo{volume}{74}}, \bibinfo{pages}{053801}
  (\bibinfo{year}{2006}).

\bibitem[{\citenamefont{Champenois et~al.}(2004)\citenamefont{Champenois,
  Houssin, Lisowski, Hagel, Knoop, Vedel, and Vedel}}]{champenois04}
\bibinfo{author}{\bibfnamefont{C.}~\bibnamefont{Champenois}},
  \bibinfo{author}{\bibfnamefont{M.}~\bibnamefont{Houssin}},
  \bibinfo{author}{\bibfnamefont{C.}~\bibnamefont{Lisowski}},
  \bibinfo{author}{\bibfnamefont{G.}~\bibnamefont{Hagel}},
  \bibinfo{author}{\bibfnamefont{M.}~\bibnamefont{Knoop}},
  \bibinfo{author}{\bibfnamefont{M.}~\bibnamefont{Vedel}}, \bibnamefont{and}
  \bibinfo{author}{\bibfnamefont{F.}~\bibnamefont{Vedel}},
  \bibinfo{journal}{Phys. Lett. A} \textbf{\bibinfo{volume}{331}},
  \bibinfo{pages}{298} (\bibinfo{year}{2004}).

\bibitem[{\citenamefont{Houssin et~al.}(2003)\citenamefont{Houssin, Courteille,
  Champenois, Herbane, Knoop, Vedel, and Vedel}}]{houssin03}
\bibinfo{author}{\bibfnamefont{M.}~\bibnamefont{Houssin}},
  \bibinfo{author}{\bibfnamefont{P.}~\bibnamefont{Courteille}},
  \bibinfo{author}{\bibfnamefont{C.}~\bibnamefont{Champenois}},
  \bibinfo{author}{\bibfnamefont{M.}~\bibnamefont{Herbane}},
  \bibinfo{author}{\bibfnamefont{M.}~\bibnamefont{Knoop}},
  \bibinfo{author}{\bibfnamefont{M.}~\bibnamefont{Vedel}}, \bibnamefont{and}
  \bibinfo{author}{\bibfnamefont{F.}~\bibnamefont{Vedel}},
  \bibinfo{journal}{Appl. Optics} \textbf{\bibinfo{volume}{42}},
  \bibinfo{pages}{4871} (\bibinfo{year}{2003}).

\bibitem[{\citenamefont{Hagel et~al.}(2005)\citenamefont{Hagel, Houssin, Knoop,
  Champenois, Vedel, and Vedel}}]{hagel05}
\bibinfo{author}{\bibfnamefont{G.}~\bibnamefont{Hagel}},
  \bibinfo{author}{\bibfnamefont{M.}~\bibnamefont{Houssin}},
  \bibinfo{author}{\bibfnamefont{M.}~\bibnamefont{Knoop}},
  \bibinfo{author}{\bibfnamefont{C.}~\bibnamefont{Champenois}},
  \bibinfo{author}{\bibfnamefont{M.}~\bibnamefont{Vedel}}, \bibnamefont{and}
  \bibinfo{author}{\bibfnamefont{F.}~\bibnamefont{Vedel}},
  \bibinfo{journal}{Review of Scientific Instruments}
  \textbf{\bibinfo{volume}{76}}, \bibinfo{pages}{123101}
  (\bibinfo{year}{2005}).

\bibitem[{\citenamefont{Drever et~al.}(1983)\citenamefont{Drever, Hall,
  Kowalski, Hough, Ford, Munley, and Ward}}]{drever83}
\bibinfo{author}{\bibfnamefont{R.}~\bibnamefont{Drever}},
  \bibinfo{author}{\bibfnamefont{J.}~\bibnamefont{Hall}},
  \bibinfo{author}{\bibfnamefont{F.}~\bibnamefont{Kowalski}},
  \bibinfo{author}{\bibfnamefont{J.}~\bibnamefont{Hough}},
  \bibinfo{author}{\bibfnamefont{G.}~\bibnamefont{Ford}},
  \bibinfo{author}{\bibfnamefont{A.}~\bibnamefont{Munley}}, \bibnamefont{and}
  \bibinfo{author}{\bibfnamefont{H.}~\bibnamefont{Ward}},
  \bibinfo{journal}{Appl. Phys. B} \textbf{\bibinfo{volume}{31}},
  \bibinfo{pages}{97} (\bibinfo{year}{1983}).

\bibitem[{\citenamefont{Numata et~al.}(2004)\citenamefont{Numata, Kemery, and
  Camp}}]{numata04}
\bibinfo{author}{\bibfnamefont{K.}~\bibnamefont{Numata}},
  \bibinfo{author}{\bibfnamefont{A.}~\bibnamefont{Kemery}}, \bibnamefont{and}
  \bibinfo{author}{\bibfnamefont{J.}~\bibnamefont{Camp}},
  \bibinfo{journal}{Phys. Rev. Lett.} \textbf{\bibinfo{volume}{93}},
  \bibinfo{pages}{250602} (\bibinfo{year}{2004}).

\bibitem[{sol()}]{solidworks08}
\bibinfo{note}{Dassault Systèmes SolidWorks Corp.}

\bibitem[{cor()}]{corning03}
\bibinfo{note}{Corning technical notice}.

\bibitem[{\citenamefont{Bergquist et~al.}(1994)\citenamefont{Bergquist, Itano,
  and Wineland}}]{bergquist94}
\bibinfo{author}{\bibfnamefont{J.}~\bibnamefont{Bergquist}},
  \bibinfo{author}{\bibfnamefont{W.}~\bibnamefont{Itano}}, \bibnamefont{and}
  \bibinfo{author}{\bibfnamefont{D.}~\bibnamefont{Wineland}}, in
  \emph{\bibinfo{booktitle}{Frontiers in Laser Spectroscopy}}, edited by
  \bibinfo{editor}{\bibfnamefont{T.~W.} \bibnamefont{H{\"a}nsch}}
  \bibnamefont{and} \bibinfo{editor}{\bibfnamefont{M.}~\bibnamefont{Inguscio}},
  \bibinfo{organization}{Proceedings of the Int. School of Physics "Enrico
  Fermi"} (\bibinfo{publisher}{North Holland}, \bibinfo{address}{Amsterdam},
  \bibinfo{year}{1994}), vol. \bibinfo{volume}{120}, pp.
  \bibinfo{pages}{359--376}.

\bibitem[{\citenamefont{Halford}(1971)}]{halford71}
\bibinfo{author}{\bibfnamefont{D.}~\bibnamefont{Halford}},
  \bibinfo{journal}{Proceedings Frequency Standards and Metrology Seminar,
  Laval University, Quebec, Canada} pp. \bibinfo{pages}{431--466}
  (\bibinfo{year}{1971}).

\bibitem[{\citenamefont{Alnis et~al.}(2008)\citenamefont{Alnis, Matveev,
  Kolachevsky, Udem, and H\"{a}nsch}}]{alnis08}
\bibinfo{author}{\bibfnamefont{J.}~\bibnamefont{Alnis}},
  \bibinfo{author}{\bibfnamefont{A.}~\bibnamefont{Matveev}},
  \bibinfo{author}{\bibfnamefont{N.}~\bibnamefont{Kolachevsky}},
  \bibinfo{author}{\bibfnamefont{T.}~\bibnamefont{Udem}}, \bibnamefont{and}
  \bibinfo{author}{\bibfnamefont{T.~W.} \bibnamefont{H\"{a}nsch}},
  \bibinfo{journal}{Physical Review A (Atomic, Molecular, and Optical Physics)}
  \textbf{\bibinfo{volume}{77}}, \bibinfo{eid}{053809}
  (pages~\bibinfo{numpages}{9}) (\bibinfo{year}{2008}),
  \urlprefix\url{http://link.aps.org/abstract/PRA/v77/e053809}.

\bibitem[{\citenamefont{Dubé et~al.}(2009)\citenamefont{Dubé, Madej, Bernard,
  Marmet, and Shiner}}]{dube09}
\bibinfo{author}{\bibfnamefont{P.}~\bibnamefont{Dubé}},
  \bibinfo{author}{\bibfnamefont{A.~A.} \bibnamefont{Madej}},
  \bibinfo{author}{\bibfnamefont{J.~E.} \bibnamefont{Bernard}},
  \bibinfo{author}{\bibfnamefont{L.}~\bibnamefont{Marmet}}, \bibnamefont{and}
  \bibinfo{author}{\bibfnamefont{A.~D.} \bibnamefont{Shiner}},
  \bibinfo{journal}{Appl. Phys. B} \textbf{\bibinfo{volume}{95}},
  \bibinfo{pages}{43} (\bibinfo{year}{2009}).

\bibitem[{\citenamefont{Legero and Sterr}(2009)}]{legero09}
\bibinfo{author}{\bibfnamefont{T.}~\bibnamefont{Legero}} \bibnamefont{and}
  \bibinfo{author}{\bibfnamefont{U.}~\bibnamefont{Sterr}}
  (\bibinfo{year}{2009}).

\bibitem[{\citenamefont{Shevy et~al.}(1993)\citenamefont{Shevy, Kitching, and
  Yariv}}]{shevy93}
\bibinfo{author}{\bibfnamefont{Y.}~\bibnamefont{Shevy}},
  \bibinfo{author}{\bibfnamefont{J.}~\bibnamefont{Kitching}}, \bibnamefont{and}
  \bibinfo{author}{\bibfnamefont{A.}~\bibnamefont{Yariv}},
  \bibinfo{journal}{Opt. Lett.} \textbf{\bibinfo{volume}{18}},
  \bibinfo{pages}{1071} (\bibinfo{year}{1993}).

\end{thebibliography}

\end{document}